\documentclass[twocolumn,showpacs,preprintnumbers,amsmath,amssymb]{revtex4}
\usepackage{graphicx}
\usepackage{dcolumn}
\usepackage{bm}

\begin{document}
\title{On the electric activity of superfluid systems}
\author{S. I. Shevchenko}
\author{A. S. Rukin}
\affiliation{B. I. Verkin Institute for Low Temperature Physics
and Engineering, Kharkov, Ukraine}
\email{shevchenko@ilt.kharkov.ua}
\pacs{67.90.+z}
\begin{abstract}
The Keldysh's theory of superfluidity of rarefied electron-hole
gas is generalized to a case of possible pair polarizability. It
was shown that inhomogeneity of the system leads to dipole moment
which is proportional to the density gradient. The dipole moment
appears also near boundaries of the system. It was determined that
quantized vortices in a magnetic field carry a real electric
charge. In He II at $H=10$ T and helium rotation velocity
$10^2$\,s$^{-1}$ the charge density is about $10^4e$\,cm$^{-3}$,
where $e$ is the electron charge.
\end{abstract}
\maketitle

Recent experiments \cite{1} -- \cite{4} revealed that motion of
superfluid helium is accompanied by appearance of electric fields
in the system, and the effect disappears at the transition to the
normal state. These experiments attracted a great interest and
stimulated a number of theoretical investigations \cite{5} --
\cite{9} in which attempts were made to explain these phenomena.
However, the physical nature of the effects observed is not
understood until the present. In this situation it seems to us
that for a qualitative explanation of the electric effects in a
moving superfluid system it is reasonable to proceed from a simple
model which allows a consistent microscopical solution. Our theory
may be considered as a purely model one. But it can also
quantitatively describe two real physical systems: spin polarized
atomic hydrogen and electron-hole gas (exciton gas). Below we will
limit ourselves to a model of rarefied electron-hole gas in which
the size of a pair is much less than the average distance between
the pairs. The properties of this gas were studied by Keldysh
\cite{10}, who showed that the behavior of this system in the
superfluid state can be described by a complex order parameter
$\Phi({\bf r}_1,{\bf r}_2)$, which has a meaning of the pair's
wave function. Ignoring the pair's internal structure (i. e.
assuming  ${\bf r}_1\hm={\bf r}_2$), Keldysh obtained an equation
for $\Phi$ which coincides with a Gross - Pitaevskij equation. As
far as the effects interesting us can appear only if the electron
and hole coordinates do not coincide, it is necessary to find the
function $\Phi({\bf r}_1,{\bf r}_2)$ at ${\bf r}_1\hm\neq{\bf
r}_2$. This problem is solved in our work. We obtain an equation
for  $\Phi$ at ${\bf r}_1\hm\neq{\bf r}_2$. It turns out to be a
nonlinear integro-differential equation which can be solved if the
order parameter varies slowly at lengths of the order of pair's
size. Using the obtained $\Phi({\bf r}_1,{\bf r}_2)$ we can find
the superfluid density $\rho_s(\bf r)$, the superfluid velocity
${\bf v}_s(\bf r)$ and the density of the system's dipole moment
${\bf P}(\bf r)$.

Let's proceed to the solution. We will consider electron-hole gas
with particles which interact according to the Coulomb law. Due to
the attraction between electrons and holes they form bound pairs,
and in the spatially homogeneous case the ground state wave
function coincides with the BCS function. However, we are
interested in the spatially inhomogeneous case, when the density
and the velocity of pairs can be functions of coordinates and
time. Keldysh \cite{10} argued that in this case the wave function
$|\Phi\rangle$ can be looked for in the form
\begin{multline}
| \Phi  \rangle  = \exp \biggl\{ -{\int {\Bigl[ {\Psi _h^ +
({\bf{r}_1})\Phi ({\bf{r}_1},{\bf{r}_2})e^{\frac{i}{\hbar }\mu t}
\Psi _e^ + ({\bf{r}_2})}}} -\\- {{{h.c.}\Bigr] }
d{\bf{r}_1}d{\bf{r}_2}} \biggr\}| 0 \rangle.
\end{multline}
Here  $\Psi _e ({\bf{r}})$ and $\Psi _h  ({\bf{r}})$  are Fermi
operators of destruction of electrons and holes correspondingly at
a point $\bf r$. It is useful to mention that in the spatially
homogeneous case Keldysh's wave function transforms to the BCS
function, in which $u_{\bf{k}}  = \cos \Phi ({\bf{k}})$,
$v_{\bf{k}}  = \sin \Phi ({\bf{k}})$, where  $\Phi ({\bf{k}})$ are
Fourier components of $\Phi({\bf r}_1,{\bf r}_2)$. The unknown
function $\Phi({\bf r}_1,{\bf r}_2)$ can be found varying the
system's energy (more exactly, the functional $E\{ {\Phi ^* ,\Phi
} \} $) by $\Phi$. Calculating the functional $E\{ {\Phi ^* ,\Phi
} \}$ in the low density limit, varying the difference $E-\mu N$,
where $N$ is the total number of particles, and equating the
result to zero, we obtain an equation ($\Phi^+({\bf r},{\bf
r'})=\Phi^*({\bf r'},{\bf r})$):
\begin{multline}
 \left( { - \frac{{\hbar ^2 }}{{2m_h }}\frac{{\partial ^2 }}
 {{\partial {\bf{r}}_1^2 }} - \frac{{\hbar ^2 }}{{2m_e }}\frac{{\partial ^2 }}
 {{\partial {\bf{r}}_2^2 }} - \frac{{e^2 }}{{\left| {{\bf{r}}_1  - {\bf{r}}_2 }
  \right|}}} \right)\Phi ({\bf{r}}_1 ,{\bf{r}}_2 ) +  \\
  + \int {R({\bf{r}}_1 ,{\bf{r}}_2 ,{\bf{r}}_3 ,{\bf{r}}_4 )\Phi ({\bf{r}}_1 ,{\bf{r}}_3 )
  \Phi ^ +  ({\bf{r}}_3 ,{\bf{r}}_4 )}\times\\\times \Phi ({\bf{r}}_4 ,{\bf{r}}_2 )
  d{\bf{r}}_3 d{\bf{r}}_4 = \mu \Phi ({\bf{r}}_1 ,{\bf{r}}_2 ).
  \label{main}
 \end{multline}
Here the kernel $R$ equals to
\begin{multline}
R= e^2 \left[ {\frac{1}{2}
     \left( {\frac{1}{{\left| {{\bf{r}}_1  - {\bf{r}}_3 }\right|}} +
     \frac{1}{{\left| {{\bf{r}}_4  - {\bf{r}}_2 } \right|}}+\frac{1}{{\left| {{\bf{r}}_4  - {\bf{r}}_3 } \right|}} }
     \right.}\right.+\\
     +\left.{\left.{
          \frac{1}{{\left| {{\bf{r}}_1  - {\bf{r}}_2 } \right|}}} \right) -
     \left( {\frac{1}{{\left| {{\bf{r}}_1  - {\bf{r}}_4 } \right|}} +
     \frac{1}{{\left| {{\bf{r}}_2  - {\bf{r}}_3 } \right|}}} \right)}
     \right].
\end{multline}
Constructing an approximate solution, we must take into account
three characteristic energies in our problem: pair binding energy
$\epsilon_0$, energy of interaction between the pairs (per one
pair) $\epsilon_{int}={g} n$, where ${g}$ is the interaction
constant (we will calculate it below), $n$ is the density of the
pairs, and energy depending on inhomogeneity $\epsilon _{inhom}
\approx {{\hbar ^2 }}/{{2ML^2 }}$, where  $L$ is the
characteristic scale of the inhomogeneity. We will suppose that
these energies satisfy the following inequalities: $\epsilon _0
\gg \epsilon _{int}   > \epsilon _{inhom}$. The first of these
inequalities is a consequence of low density of pairs $n$ supposed
above. The second inequality means that we consider only weakly
inhomogeneous states, where the characteristic inhomogeneity scale
$L$ is greater than the coherence length $\xi \equiv ( {{{\hbar ^2
}}/{{2M{g} n}}} )^{1/2} $. As far as the energy $\epsilon_0$  is
associated with motion of the electron and the hole relatively to
each other, and other energies represent motion of the pair as a
whole, the solution of (\ref{main}) can be looked for as $ \Phi  =
[ {\phi ^{(0)} ({\bf{r}}_{12}  ) \hm+ \phi ^{(1)} ({\bf{r}}_{12}
;{\bf{R}}_{12} )}]\Psi({\bf{R}}_{12} ) $ (${\bf r}_{12}={\bf
r}_1\hm-{\bf r}_2$, ${\bf R}_{12}$ is the coordinate of the pair's
center of mass), and $\phi^{(1)}$ is $\epsilon_{int}/\epsilon_0$
times less than $\phi^{(0)}$. The $\Psi$ function (order
parameter) describes motion of the pair as a whole.

In the zero order approximation
\begin{equation}
\mu _0 \phi ^{(0)} \Psi = H_0 \phi ^{(0)} \Psi.
\end{equation}
Here $H_0$ is the Coulomb problem Hamiltonian, $\mu_0=\epsilon_0$,
 $\phi^{(0)}=\phi_0$, where $\epsilon_0$ and $\phi_0$ are correspondingly the ground state energy
and wave function.

In the next approximation
\begin{multline}
 \mu _0 \phi ^{(1)} \Psi  = H_0 \phi ^{(1)} \Psi  - \mu _1 \phi _0 \Psi   - \frac{{\hbar ^2 }}{{2M}}\frac{{\partial ^2  }}{{\partial {\bf{R}}_{12}^2 }}\phi _0 \Psi  +  \\
  + \int {R({\bf{r}}_1 ,{\bf{r}}_2 ,{\bf{r}}_3 ,{\bf{r}}_4 )\phi _0 ({\bf{r}}_{13} )\phi _0 ({\bf{r}}_{34} )\phi _0 ({\bf{r}}_{42} )}\times\\
  \times{\Psi ({\bf{R}}_{13} )\Psi ^* ({\bf{R}}_{43} )\Psi ({\bf{R}}_{42} )d{\bf{r}}_3 d{\bf{r}}_4
  }.\label{eqn-phi1}
 \end{multline}
Equation (\ref{eqn-phi1}) has a solution only if its inhomogeneous
part is orthogonal to the solution of the corresponding
homogeneous equation, i. e.  $\phi_0$. The solvability condition
leads to the following equation for  $\Psi$:
\begin{equation}
 - \frac{{\hbar ^2 }}{{2M}}\frac{{\partial ^2 \Psi }}{{\partial {\bf{R}}_{12}^2 }}
 - \mu _1 \Psi  + {g} \bigl| {\Psi ({\bf{R}}_{12} )} \bigr|^2 \Psi ({\bf{R}}_{12} ) =
 0.\label{Gross}
\end{equation}
Here the interaction constant ${g}$ equals to
\begin{multline}
{g}  \equiv \int {{\bf{R}}({\bf{r}}_1 ,{\bf{r}}_2 ,{\bf{r}}_3
,{\bf{r}}_4 )\phi _0 ({\bf{r}}_{13} )\phi _0 ({\bf{r}}_{34} )\phi
_0 ({\bf{r}}_{42} )\phi _0 ({\bf{r}}_{12}
)}\times\\\times{d{\bf{r}}_3 d{\bf{r}}_4 d{\bf{r}}_{12} }  =
\frac{{13}}{3}\pi e^2 a_B^2.
\end{multline}
While solving the equation (\ref{eqn-phi1}) we must take into
account that in the last summand, due to existence of three
$\phi_0$ functions, arguments of all $\Psi$ functions must be
close to ${\bf R}_{12}$. Solving then (\ref{eqn-phi1}) we find
that the part of $\phi^{(1)}$ which makes a nonzero contribution
to the dipole moment is equal to
\begin{equation}
 \phi ^{(1)} = \int {d{\bf r}^\prime_{12} G({\bf{r}}_{12}
,{\bf r}^\prime_{12} )} F({\bf r}^\prime_{12} ){\bf r}^\prime_{12}
\cdot \frac{{\partial \left| {\Psi _0 ({\bf{R}}_{12} )} \right|^2
}}{{\partial {\bf{R}}_{12} }}.
\end{equation}
Here $G$ is the Green function of the Coulomb problem, $ F({\bf
r}_{12} )$ is an even function of ${\bf r}_{12}$ which explicit
expression is not shown here because it is too cumbersome.

Now we can find the dipole moment density of the system.
\begin{multline}
{\bf{P}} = \int {\left[ {\phi ^{(0)*} ({\bf{r}}) + \phi ^{(1)*}
({\bf{r}})} \right]\Psi ^* ({\bf{R}})}\times \\ \times {e{\bf{r}}}
\left[ {\phi ^{(0)} ({\bf{r}}) + \phi ^{(1)} ({\bf{r}})}
\right]\Psi ({\bf{R}})d{\bf{r}}.
\end{multline}
Substituting here the $\phi_0$ and $\phi_1$ functions, we obtain
\begin{equation}
{\bf{P}} =  - A\gamma ena_B^5 \frac{{\partial n}}{{\partial
{\bf{R}}}}. \label{P}
\end{equation}
Here $\gamma=\frac{{m_h  - m_e }}{{m_h  + m_e }}$, $A$ is a
numerical coefficient, $A \hm\approx 120$. We see that an
inhomogeneity in the system leads to appearance of the dipole
moment. It becomes zero if $m_h \hm= m_e$, i. e. the dipole moment
is caused by the mass asymmetry of the pairs. Usually $m_h \hm>
m_e$. In this case in an inhomogeneous system the dipole moment's
positive end points opposite to the density gradient.

A real physical system is always limited with vessel walls.
Interaction between atoms of the system and vessel walls causes a
surface dipole moment. The dielectric constant of the
electron-hole gas is denoted by $\epsilon_1$, of the vessel walls
$\epsilon_2$. The potential of interaction between the pair and
the surface is found by means of electric image method. At large
distances to the surface $Z \hm\gg z \hm\sim a_B $ ($Z>0$) the
potential equals to
\begin{equation}
V = -\frac{1}{{16}}\frac{{\epsilon _2  - \epsilon _1 }}{{\epsilon
_2  + \epsilon _1 }}\frac{{e^2 }}{{\epsilon _1 }}\frac{1}{{Z^3
}}\left( {\rho ^2  + 2z^2 } \right)\left( {3\gamma\frac{z}{Z} + 2}
\right).\label{potential}
\end{equation}
This potential appears in the main equation (\ref{main}) and
changes both $\phi({\bf r})$ and $\Psi({\bf R})$. Solving the
equation for $\phi_0$ using the variational method, we find the
surface dipole moment, which is directed normally to the surface.
\begin{equation}
P_z  = \frac{{414}}{{37}}\gamma\frac{1}{{\epsilon _1
}}\frac{{\epsilon _2  - \epsilon _1 }}{{\epsilon _2  + \epsilon _1
}} en\frac{{a_B^5 }}{{Z^4 }}.\label{P1}
\end{equation}
We again see that the dipole moment becomes zero if $m_h \hm=
m_e$, and when $m_h \hm> m_e$ ($\gamma>0$) its positive end is
directed toward the medium with lower dielectric constant. Near a
metal surface $(\epsilon _2  \hm\to \infty )$ at $m_h \hm> m_e$
the dipole moment $P_z>0$, i. e. the electron is closer to the
surface than the hole. Near a boundary with vacuum $(\epsilon_2 =
1)$ the opposite case is realized.

The potential (\ref{potential}) influences also the order
parameter. The equation for it obtains the form
\begin{equation}
 - \frac{{\hbar ^2 }}{{2M}}\frac{{\partial ^2 \Psi }}{{\partial
 {\bf{R}}_{12}^2 }} - \mu _1 \Psi  - \frac{1}{{2\epsilon _1 }}
 \frac{{\epsilon _2  - \epsilon _1 }}{{\epsilon _2  +
 \epsilon _1 }}\frac{{e^2 a_B^2 }}{{Z^3 }}\Psi  + {g}
 \left| \Psi  \right|^2 \Psi  = 0. \label{eqn-Psi}
\end{equation}
Outside the immediate vicinity of the boundary the first summand
in (\ref{eqn-Psi}) can be omitted (Thomas -- Fermi approximation).
Consequently we find that the addition to the order parameter
caused by Van der Waals interaction with the surface, has the form
$ \left| \Psi  \right|^2  \hm= \frac{1}{{2{g} \epsilon _1
}}\frac{{\epsilon _2  - \epsilon _1 }}{{\epsilon _2  + \epsilon _1
}}\frac{{e^2 a_B^2 }}{{Z^3 }} $. This inhomogeneous addition,
according to (\ref{P}), leads to an extra dipole moment
\begin{equation}
P_z  \approx 15\gamma\frac{1}{{\epsilon _1 }}\frac{{\epsilon _2  -
\epsilon _1 }}{{\epsilon _2  + \epsilon _1 }}en\frac{{a_B^5
}}{{Z^4 }}. \label{P_additional}
\end{equation}
This expression must be added to (\ref{P1}).

To understand more clearly the cause of the dipole moment in the
multiparticle system considered, it is useful to utilize the
results for a system consisting only of two identical atoms. The
Van der Waals interaction between the atoms leads to the shift of
the electron density distribution center of each atom towards the
opposite atom, and dipole moments appear (see e. g. \cite{11}).
\begin{equation} p_1=({\bf p}_1\cdot {\bf n})=-p_2=-({\bf
p}_2\cdot {\bf n})=\frac{D}{R_{12}^7}. \label{p_1}
\end{equation}
Here ${\bf n}={\bf R}_{12}/R_{12}$. In particular, for two helium
atoms \cite{11} $D=18.4ea_B^8$.

In a rarefied system the dipole moment of an atom is formed by
adding the dipole moments which appear in this atom as a result of
interaction with all other atoms. As follows from (\ref{p_1}), in
a hypothetical medium where at $Z>0$ the atomic density equals to
$n_1$ and at $Z<0$ it equals to $n_2$, the $z$ component of the
dipole moment of an atom at the point $(0,0,Z_0)$ $(Z_0>0)$ equals
to
\begin{equation}
p_z(Z_0)=-D\int\frac
{[n_1\theta(Z)+n_2\theta(-Z)](Z-Z_0)}{[\rho^2+(Z-Z_0)^2]^4}d^2\rho
dZ.\label{PZint}
\end{equation}
Here $\theta(Z)$ is the Heaviside step function. When writing this
expression we took into account that the $z$ component of the
dipole moment of a pair of atoms is obtained multiplying
(\ref{p_1}) by cosine of the angle between the $z$ axis and the
vector connecting these atoms. As far as the distance between the
atoms cannot be less than the size of an atom, $\rho$ and $Z$ must
be cut off at the lower limit at the Bohr radius $a_B$. Further we
assume $Z_0\gg a_B$. After a simple integration in (\ref{PZint})
we obtain
\begin{equation}
p_z=\frac{\pi}{12}D\frac{n_2-n_1}{Z_0^4}. \label{p_D}
\end{equation}
If we take into consideration that for a rarefied system
$\epsilon_{1,2}\hm=1\hm+4\pi\alpha n_{1,2}$ and the polarizability
$\alpha\hm=Ca_B^3$, where $C$ is a numerical coefficient (for
hydrogen $C=\frac 92$, for helium $C=\frac 94$), it is easy to
prove that (\ref{P_additional}) coincides in letters with
(\ref{p_D}) multiplied by $n$ to obtain the dipole moment density
$\bf P$.

Above we assumed the densities $n_{1,2}$ to be homogeneous. But
near the vessel walls due to the Van der Waals interaction the
density acquires an addition $\sim\frac 1{Z^3}$, where $Z$ is the
distance from the wall. Supposing that $n_1$ is a slow function of
$Z$ and writing it as $n_1\hm=n_{10}\hm+\frac{\partial n}{\partial
Z}Z$, after the integration in (\ref{PZint}) we obtain the
addition to $p_z$
\begin{equation}
\Delta p_z=-\frac{\pi}{6a_B^3}D\frac{\partial n}{\partial Z}.
\end{equation}
After multiplying by $n$ this expression coincides in letters with
(\ref{P}). Thus, knowing the electron density distribution in a
system of two atoms (\ref{p_1}), we can qualitatively reproduce
the results of the microscopical calculation.

Only stationary states were considered until now. A more general
state is the Keldysh wave function where $\Phi({\bf R}_1,{\bf
R}_2)$ is replaced by $\Phi({\bf R}_1,{\bf R}_2,t)$, and its
dependence on $t$ is slow compared to $\exp(\frac i{\hbar} \mu
t)$. As the result, right hand sides of equations (\ref{eqn-phi1})
and (\ref{Gross}) will acquire an extra summand
$i\hbar\frac{\partial\Psi}{\partial t}$.

It is interesting to find out the relationship between obtained
results and an affirmation stated by Melnikovsky \cite{6} that
acceleration of dielectric leads to its polarization, and
\begin{equation}
{\bf P}=-\frac{\epsilon-1}{4\pi}\frac 1{2Ze}M\frac{\partial {\bf
v}}{\partial t}. \label{meln}
\end{equation}
Here (and only here) $Z$ is the atomic number. Below $Z=1$.
Denoting as $U({\bf R})$ the interaction energy of the atom in the
point ${\bf R}$ with all other atoms and taking into account again
that for a rarefied medium $\epsilon=1+4\pi n \alpha$, we obtain
from (\ref{meln})
\begin{equation}
{\bf P}({\bf R})=-\frac{\alpha n({\bf
R})}{2e}\left(-\frac{\partial U}{\partial {\bf
R}}\right).\label{meln1}
\end{equation}
If (not very justified) we spread this expression to two atoms at
points ${\bf R}$ and ${\bf R'}$, then at $|{\bf R}-{\bf R'}|\hm\gg
a_B$ the Van der Waals attraction forces will act between these
atoms and $U=-C/|{\bf R}-{\bf R'}|^6$, where $C\approx e^2 a_B^5$.
As far as ${\bf P}=n{\bf p}$, (\ref{meln1}) yields an expression
for the dipole moments of a pair of atoms which coincides by sign
and by order of magnitude with (\ref{p_1}). This result confirms
applicability of (\ref{meln}) down to atomic length scales.

In the case of a dielectric medium the total dipole moment of the
system equals to
\begin{equation}
\int{\bf P}dV=\frac{\alpha}{2e}\left[\int nUd{\bf S}-\int
U\frac{\partial n}{\partial{\bf R}}dV\right]. \label{21}
\end{equation}
We have used the expression (\ref{meln1}) and integrated its right
hand side by parts. As it was shown above, the pair interaction
energy (per one pair) equals to $gn$. Using (\ref{21}) and taking
into account $U=gn$ we obtain the volume part of the dipole moment
\begin{equation}
{\bf P}=-\frac{13}{6}\pi\alpha e a_B^2 n\frac{\partial
n}{\partial{\bf R}}.
\end{equation}
This expression coincides in letters with (\ref{P}) when
$m_h\hm\gg m_e$. The last inequality is implicitly supposed to be
fulfilled in (\ref{meln}).

In the general case the density can be a complicated function of
coordinates and time. But a superfluid system can possess
``characteristic configurations'' which are of special interest,
in particular, rectilinear vortices and vortex rings. In the case
of a rectilinear vortex the density becomes a function of distance
$\rho$ to the vortex axis. As the result a radially directed
dipole moment appears in the system.
\begin{equation}
{\bf P}=-\frac{3A}{13\pi}\frac
{na_B^3}e\frac{\hbar^2}{M\rho^3}\frac{\boldsymbol {\rho}}{\rho}.
\end{equation}
This result (with an opposite sign) was first obtained by Natsik
\cite{7} phenomenologically.

A more complicated problem about the dipole moment of a vortex
ring can be solved if we take into account that far from the
vortex core a relation is fulfilled: $ \nabla( {Mv^2}/2)\hm=-{g}
\nabla n \label{n-v} $. As far as the velocity field of a vortex
ring is known \cite{11A}, it is easy to find the dipole moment of
the ring as a function of distance to its axis. We can also find
the total dipole moment of the ring and show that it is zero
unlike the affirmations found in literature that it is nonzero.

Until now we assumed that there is no external magnetic field.
Presence of a magnetic field leads to a cardinal change in the
situation. The effect is caused by appearance of a summand $\frac
e{c}({\bf v}\times{\bf H})\hm\cdot{\bf r}_{12}$ in the
Hamiltonian. Here $\bf v=\frac{\hbar}M\nabla\phi$ is the velocity
of the pairs. The system's response to this addition is similar to
the response to an external electric field $E_{eff}\hm=\frac 1c
({\bf v}\times{\bf H})$. Like a real electric field $E_{eff}$
leads to appearance of a dipole moment in the system:
\begin{equation}
{\bf P}=\frac{\alpha}{c}({\bf v}\times{\bf H})n,
\label{P_magnetic}
\end{equation}
where $\alpha$ is the pair polarizability. This dipole moment is
connected with the Lorentz force acting on the positive and
negative charges in opposite directions. If we substitute the
velocity field of a rectilinear vortex $\bf v$ in
(\ref{P_magnetic}) we will see that a radial dipole moment, as
without a magnetic field, will appear around the vortex axis. But
it is extremely important that this dipole moment, similarly to
the velocity field, decreases with distance from the axis as
$1/\rho$. As the result the total surface polarization charge
$\int {\bf P}\cdot d{\bf S}$ does not depend on $\rho$ and on the
surface shape if the surface remains cylindrical. Due to electric
neutrality of the system this result leads that the vortex core
possesses the opposite sign charge which equals per unit length
($\bf H$ being parallel to the cylinder axis)
\begin{equation}
q=\pm\frac{\alpha\hbar}{Mc} 2\pi Hn.
\end{equation}
The charge sign depends on the sign of vortex circulation. If the
vessel with helium rotates with frequency $\Omega$, the density of
the vortices equals to $n_v=M\Omega/\pi\hbar$, and the total
charge of the vortices per unit volume is
\begin{equation}
Q=\pm\frac{2\alpha n}{c}H\Omega.
\end{equation}
For He II in a magnetic field 10\,T and at rotation speed
$10^2$\,s$^{-1}$ the charge density is approximately
$10^4e$\,cm$^{-3}$, where $e$ is the electron charge.

Appearance of electric charge in the vortex core will lead to
nonzero dipole moment of vortex-antivortex pairs (in a 2D system)
and vortex rings (in a 3D system).

We must mention that electric polarization and electric charge in
the vortex cores in $^3$He were considered in \cite{12}. Unlike
our work, the polarization and the charge of the vortices in
$^3$He are caused by flexoelectric effect.

Now let us show that due to existence of the dipole moment $\bf P$
near the metal boundary a second sound wave incident to it will
lead to an oscillating potential difference between the metal and
the helium. Since the induction $ {\bf{D}} \hm= {\bf{E}} \hm+ 4\pi
{\bf{P}}$ must be continuous and in metal ${\bf D} \hm=0$, in
helium filling the metal vessel an electric field appears in the
vicinity of the vessel walls: ${\bf E}\hm=-4\pi{\bf P}$. This
field, similar to the dipole moment $\bf P$, is constant at
constant density $n$. But when a second sound wave propagates,the
density $n$ and the dielectric constant $\epsilon$ oscillate in
time, more exactly, they obtain oscillating additions (see below).
As far as ${\bf D}\hm=\epsilon{\bf E}$ and $\delta {\bf{D}} \hm=
\delta \epsilon   {\bf{E}} \hm+ \epsilon  \delta {\bf{E}} = 0$,
the oscillating addition $\delta \epsilon$ leads to similar
addition to the field $\delta {\bf{E}} \hm=  - \frac{{\delta
\epsilon }}{\epsilon }{\bf{E}} $. This addition is the cause of
the potential difference between metal and helium.

The oscillating addition $\delta\epsilon$ can be found taking into
account that the Clausius -- Mossotti relation $\frac{{\epsilon  -
1}}{{\epsilon  + 2}} \hm= 4\pi n\alpha$ leads to $ \delta \epsilon
(t) = \frac{1}{3}(\epsilon  - 1)(\epsilon  +
2)\frac{1}{n}\frac{{\partial n}}{{\partial T}}\delta T(t)$.

The expansion coefficient of liquid helium $ \beta  \equiv
\frac{1}{n}\frac{{\partial n}}{{\partial T}}$ can be taken from
the experiments (see \cite{13}). For example, $ \beta  \hm=  - 0.1
\cdot 10^{ - 3} $\,K$^{-1} $ at $ T \hm= 1.138$\,K and $ \beta
\hm= - 26.34 \cdot 10^{ - 3} $\,K$^{-1}$ at $ T \hm= 2.15$\,K.
Since $\epsilon  - 1 = 5.7 \cdot 10^{ - 2} $, at $T = 2.15$\,K the
oscillating addition equals to $\delta \epsilon  =  - 1.5 \cdot
10^{ - 3} \delta T$ where $\delta T$ is in kelvins. The constant
field at the surface of helium substantially depends on $z_0$.
Choosing $z_0=5\cdot10^8$\,cm, we obtain $E\sim 10^6$\,V/cm. In
this case the oscillating potential difference between metal and
helium $\delta \phi  \cong \delta \epsilon Ez_0$ at $ T = 2.15$\,K
will be equal to $ \delta \phi \approx  - 5 \cdot 10^{ - 4} \delta
T$\,V. This value is close to the experimental one, but strongly
depends on temperature. At $ T = 1.138$\,K we obtain $ \delta \phi
\approx 10^{ - 6} \delta T$\,V, i. e. two orders less than
experimental data.

Comparing theory and experiments we must take into consideration
that without special preparing of the vessel containing helium its
walls are covered with several layers of adsorbed atoms. The
interaction of these atoms with metal will induce dipole moments
in them and electric fields. The polarizability of adsorbed atoms
is almost two orders greater than the helium polarizability,
therefore, the induced electric field in the adsorbed atoms will
be correspondingly greater. Estimates given above allow to expect
that the scale of the effect will be the same as in experiments.
As far as the superfluid transition temperature is not an
exceptional point for adsorbed atoms, there are no reasons to
expect a substantial temperature dependence of $\delta\phi$ in the
temperature interval $1.5-2$\,K in which experiments were
performed.

It should be noted that the effects predicted in this work do not
give a comprehensive explanation of the experiments \cite{1,2}. In
our opinion, for such an explanation we need a new key idea.

We express our thanks to L. A. Pastur for useful discussion of the
work. We also acknowledge A. S. Rybalko for the possibility of
learning the results of his experiments before their publishing.

\end{document}